\documentclass[final,1p,times]{elsarticle}

\usepackage{lineno,hyperref}
\modulolinenumbers[5]

\usepackage{booktabs}
\usepackage{array,caption,threeparttable}
\usepackage[font=small,labelfont=bf]{caption}
\usepackage{amsmath,amssymb,amsfonts}
\usepackage{textcomp}

\def\BibTeX{{\rm B\kern-.05em{\sc i\kern-.025em b}\kern-.08em		T\kern-.1667em\lower.7ex\hbox{E}\kern-.125emX}}

\usepackage{graphicx}
\usepackage{epsfig}
\usepackage{subfigure}

\usepackage{array}

\usepackage{multirow}

\usepackage{float}
\usepackage{caption}
\usepackage{color}
\usepackage{verbatim}
\usepackage{bbding}
\usepackage{enumerate}
\usepackage{ntheorem}
\usepackage{cases}
\usepackage{booktabs} 
\usepackage{lscape}
\usepackage{makecell}

\usepackage{bm}

\usepackage{amsfonts}

\usepackage{mathrsfs}

\newtheorem{lemma}{Lemma}
\newtheorem{theorem}{Theorem}

\newtheorem{corollary}{Corollary}

\def\be{\begin{equation}}
\def\ee{\end{equation}}
\def\bea{\begin{eqnarray}}
\def\eea{\end{eqnarray}}
\newcommand{\re}[1]{(\ref{#1})}
\def\ne{ \nonumber \\ }
\def\nn{ \nonumber }
 \def\QED{~\rule[-1pt]{5pt}{5pt}\par\medskip}

\begin{document}

\begin{frontmatter}

\title{
Constructive RNNs:
An Error-Recurrence Perspective
on Time-Variant Zero Finding Problem Solving Under Uncertainty
}



\author[mymainaddress]{Mingxuan Sun\corref{mycorrespondingauthor}}
\cortext[mycorrespondingauthor]{Corresponding author.}
\ead{mxsun@zjut.edu.cn}

\author[mymainaddress]{Xing Li}
\ead{1416440872@qq.com}


\author[mymainaddress]{Han Wang}
\ead{408687710@qq.com}

\address[mymainaddress]{College of Information Engineering, Zhejiang University of
Technology, Hangzhou, 310023, China}

\begin{abstract}
When facing time-variant problems in analog computing,
the desirable RNN design requires finite-time convergence
and robustness with respect to
various types of uncertainties,
due to the time-variant nature and difficulties in implementation.
It is very worthwhile to explore
terminal zeroing neural networks,
through examining and applying
available attracting laws.
In this paper, from a control-theoretic point of view,
an error recurrence system  approach is presented
by equipping with uncertainty compensation
in the pre-specified error dynamics,
capable of enhancing robustness properly.
Novel rectifying actions
are designed to make finite-time settling so that the convergence speed and the computing accuracy of time-variant computing can be improved.
Double-power and power-exponential rectifying actions are respectively formed to construct specific models,
while the particular expressions of settling time function for the former
are presented, and for the latter the proximate settling-time estimations are given,
with which the fixed-time convergence of the
corresponding models is in turn established.
Moreover, the uncertainty compensation
by the signum/smoothing-signum techniques are adopted for finite-duration stabilization.
Theoretical results are presented
to demonstrate effectiveness (involving
fixed-time convergence and
robustness)
of the proposed computing schemes for the time-variant QP problem solving.
\end{abstract}


\end{frontmatter}


\section{Introduction}
\label{sec.introduction}

Recurrent neural networks (RNNs)
have architecture with feedback loops, offering a
tool for online solving problems in science and engineering \cite{hop84,cpwu00,ham01}.
As an alternative to analog computing,
zeroing neural networks
(ZNNs, see \cite{zhang02,zhang05,zhang16} and the references therein),
are of RNN-like structure
aiming to the time-variant problem solving,
where with the pre-specified error dynamics,
the existence and global stability of solutions of the designed neural networks
are guaranteed.
Efficient models were particularly designed and implemented
for various time-variant problems including
matrix inversion and pseudoinverse,
linear/nonlinear matrix equations,
matrix inequalities, quadratic/nonlinear programming, etc.


The theoretical solutions of time-variant problems change with time and
the convergence performance is the key to solve them.
It is highly desirable to achieve the perfect result, namely, zero-error convergence over the entire time interval.
However,
theoretical solutions on an initial interval are indeed difficult to be obtained in the presence of initial errors.
It is realistic
for an RNN in realtime computation
that a solution is available after certain initial interval.
Fortunately,
the way for RNN designs to get out of the difficult situation
is to apply the finite-time stability theory.
Finite-time convergent ZNN models enable to provide accurate solutions
after the settling time.
As a supplement to most
of existing results guarantee asymptotical stability and exponential stability,
the finite-time convergent computing methods
have received increased attention,
and
early works were found in \cite{cheng09,liu11,li13}.
The signum unit was employed in \cite{cheng09}
for switching the neural network structure;
the hard limit activation function was used in \cite{liu11}; and the sign-bi-power one was suggested in \cite{li13}.
The extensive and comprehensive studies on ZNN models are found in \cite{
shen15,xiao15,sun15,jin17nc,xiao23},
considering the fact that real-time performance is highly demanded for time-variant problems in practice.
Very recently,
various types of activation functions have been summarized in \cite{sun21a}, where the asymptotically/finite-time convergent ZNN models
were examined for characterizing the error evolution and its attractiveness.


For the time-variant problem solving,
one would like to achieve
the prescribed convergence performance,
which could dramatically
improve the computing accuracy.
Fixed-time stability is a contemporary concept
assuring an upper bound on the settling time function which is independent of initial conditions.
The concept of predefined-time stability is so helpful that the settling time is pre-specified and adjustable.
Giving an estimation of the bound on the settling-time function of the system undertaken,
the system can be re-constructed to realize
predefined time stability with
predefined settling time proportional to the inverse of the estimation.
We refer interested readers to the recent literature \cite{polyakov12,efimov21} and the references therein.
Both stability concepts are especially useful
to the time-variant computing problem solving,
because the exact solution can be obtained definitely after the pre-specified instant.
However,
the estimations for settling time reported
in many related works are conservative,
due to the estimate being much larger than the actual one typically for the double-power systems.
The closed-form expressions of settling time functions of the typical nonlinear systems,
among others,
were presented by means of special functions in \cite{hu22,sun21b,sun23},
and the tight bound on settling time
was given for exact estimation.
In \cite{sun20}, the closed-form
settling time function of
two-phase systems was obtained,
which still admits fixed-time attractors.
It is also desirable for neural network designs that achieve the fixed-/predefined-time convergence.
The reported ZNNs
 can realize such computing performance so that theoretical solutions can be achieved in a fixed time
\cite{jin17nc,xiao23,sun24,li24}.
Many existing works focus on constructing the error dynamics and its convergence analysis, especially for speeding up the convergence in the absence of uncertainty.


The robustness of neural networks
is crucial for analog computing,
since external disturbances and noises inevitably exist in the implementation.
Varying-parameter ZNNs were proposed in \cite{zhang18}, and it was show that by monotonically increasing the parameters,
these ZNNs are of super exponential convergence and the residual errors converge to zero even under perturbation situations.
In \cite{zhang19},
Power-type varying-parameters were introduced in
forming the error dynamics
to address the robustness issue,
by which
the computing error caused by possible differentiation
and implementation errors can be made arbitrarily small through simply increasing the design parameters.
Integral-enhanced computing schemes have been shown
to possess robustness and achieve the tolerance of constant disturbances,
for solving matrix inversion
and
constrained nonlinear
optimization
\cite{jin16b,li20,wei22}.
By simultaneously considering
finite-time convergence and robustness,
in \cite{xiao19a},
PID-type error dynamics
was constructed for solving Lyapunov equation
in the presence of various kinds of additive noises.
In \cite{li20a,dai21,zhangM22},
both linear and signum terms
of the designed error dynamics
were shown to be effective in
handling bounded nonvanishing noises.
From the above mentioned, a control-theoretic approach is expected
to contribute to a sound design rationale for enhancing robustness of ZNN models, not only aiming at specific types of uncertainties.

The problem to find a zero of a given time-invariant function was addressed from the point of view of feedback control,
in \cite{bhaya07},
showing how to arrive at a control taxonomy of the methods to solve computing problems.
Exactly speaking, the computing problem undertaken can be expressed as the regulation problem in control.
The proposed approach offers
a unified framework to derive zero finding algorithms (including
the Newton-Rapshon algorithm).
In addition, a new interpretation of the conjugate gradient algorithm as a proportional-derivative controller was made.
In \cite{noroozi09}, the design
method is extended to the case when the Jacobian can be decomposed into a known part and a partially known part.
The
extension of existing zero finding methods to this setting allows
the consideration of functions which have singular Jacobians as well as the underdetermined case.
Continuous effort has been made to
extend this method
for solving time-variant zero finding problems in \cite{jin17,jin18,qi22}.
An integration term was introduced into the error model, in \cite{jin17},
and a generalized proportional integral-derivative controller was constructed for the problem solving.
In \cite{jin18},
the proposed RNN model remedies limitations of the activated functions, through the removal of the convex restriction.
In \cite{qi22},
discrete recurrent neural dynamics were constructed to robustly cope with noise,
showing how iterative methods for solving time-variant computing problems can be used
in a control framework.

By building into the constructive error dynamics of ZNNs a certain of flexibility for zero finding problem solving,
in this paper, an error recurrence system (ERS) approach is proposed,
which is capable of enhancing robustness properly with respect to various uncertainties in implementation.
A rectifying action,
aiming to improve convergence rate,
is involved in the error dynamics and
with the chosen input action,
the computing scheme can be designed with ease.
Double-power and power-exponential rectifying actions are applied, respectively, with which specific forms of ZNN models are constructed.
The uncertainty compensation
by the signum/smoothing-signum techniques are adopted for achieving finite-duration stabilization,
and the constructed models
are analyzed for solving QP problems in the presence of uncertainty.

\section{TZNNs as systems with error recurrence
} \label{sec.res}

Let us begin to consider time-variant zero finding problem, which is generally
to find the vector-valued variable,
\[
\mathbf{x}^{*}(t)\in \mathcal{R}^{n}, ~~{\rm for}~~ t \in [0,\infty)
\]
such that
\[
\mathbf{f}(\mathbf{x}^{*}(t), t) = 0
\]
with
$\mathbf{f}: \mathcal{R}^n \times [0,+\infty)\rightarrow \mathcal{R}^n$,
 a vector-valued continuous time-variant nonlinear function.
Here, we suppose the existence of unique solution
$\mathbf{x}^{*}(t)$, for $t \in [0,\infty)$.
Let us denote by
\[
\mathbf{e}(t) = 0 - \mathbf{f}(\mathbf{x}(t), t)
\]
the computing error.

With the definition of the computing error, the problem of
zero finding can be formulated in the following ways:

{\bf Convergence Problem (CP)}

Find the solution by assuring the convergence of $\mathbf{e}(t)$,
in the absence of disturbances.

{\bf Robust Convergence Problem (RCP)}

Find the solution, in the presence of disturbances,
by making that $\mathbf{e}(t)$ converges to a neighborhood of the origin
and the neighborhood radius is proportional to the upper bound on disturbances.

\begin{figure} [htbp]
  \centering
{
  \begin{minipage}{6cm}
  \includegraphics[width=6cm,height=3.5cm]{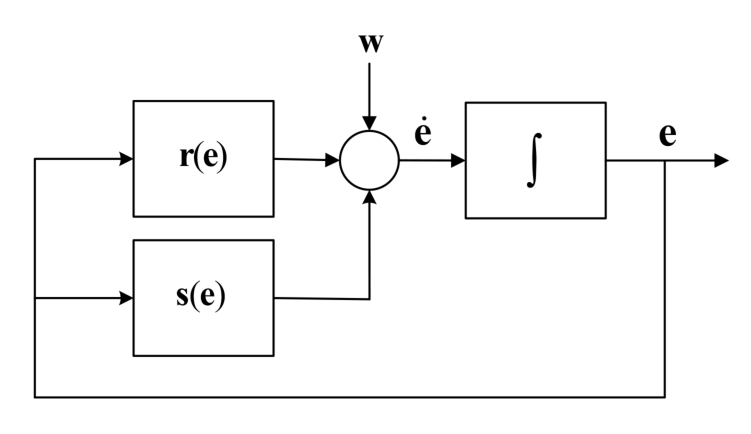}
  \end{minipage}
  }
  \caption{
 Error recurrence system}
 \label{fig.res}
  \end{figure}

From the perspective of error recurrence system (ERS) for the problem solving,
in this paper, we address the RCP that can be solved through
directly developing ZNN models.
Here, for the problem solving under uncertainty, the formulation of ERS is presented by systemizing ZNN models,
described by the following differential equation:
\bea
\label{err.dyn}
\left \{
\begin{array}{l}
 \dot {\mathbf{e}} = \mathbf{r} + \mathbf{s} + \mathbf{w} \\
 \mathbf{r} = \mathbf{r}(\mathbf{e}) \\
 \mathbf{s} = \mathbf{s}(\mathbf{e})
 \end{array}
\right.
\eea
where
$\mathbf{r}$ is the rectifying action,
$\mathbf{s}$ is the compensation term, and
$\mathbf{w}$ is the lumped disturbances that
exist
when implementing the computing scheme.
In fact,
Eq. \re{err.dyn} indicates a feedback loop
with the error ${\bm e}$ being not only the output but also the input,
and
$\mathbf{u} =  \mathbf{r} + \mathbf{s}$ represents the control signal,
as seen in Fig. \ref{fig.res},
illustrating that our approach to problem solving is in a recurrent manner.

The rectifying action $\mathbf{r}$ is constructed for assuring the convergence performance
of system \re{err.dyn}, in the absence of $\mathbf{w}$.
Once
$\mathbf{e}$ converges to zero,
$\mathbf{x}$ is driven accordingly to the solution $\mathbf{x}^{*}$.
However, in the implementation,
we have to take into account the presence of $\mathbf{w}$,
and the computing error $\mathbf{e}$
governed by \re{err.dyn}
obviously depends on $\mathbf{w}$.
The term $\mathbf{s}$ is employed
to improve the computing performance via rejecting $\mathbf{w}$.
Then, we summarize the design procedure as follows:

i. Design the rectifying action $\mathbf{r}$, in the absence of $\mathbf{w}$, by applying an appropriate attracting law
to guarantee convergence of $\mathbf{e}$; and

ii. Redesign the compensation $\mathbf{s}$, in the presence of $\mathbf{w}$,
for the purpose of robustness performance improvement through an efficient compensation technique.

For constructing the specific ${\bm r}$,
we would like to provide novel designs
associated with more close estimates for the settling time.
These rectifying actions presented are on the basis of the recently developed finite-time system theory for computing.
Then we would like to give a redesign
for choosing the action $\bm{s}$.

The ERS approach is applicable to computational tasks such as time-variant optimization problem  solving.
As one typical example,
let us consider the time-variant
QP in the form of
\bea   \label{fx1}
\left\{
\begin{array}{ll}
\min\limits_{x(t)}&\frac{1}{2}x^{T}(t)G(t)x(t)+c^{T}(t)x(t)  \\
s.t. & A(t)x(t)=b(t)
 \end{array}
\right.
\eea
where $t \in [0,\infty)$,
$x(t) \in \mathcal{R}^{n}$
represents the unknown vector to be found,
$G(t) \in \mathcal{R}^{n \times n}$ is a symmetric positive-definite matrix,
$A(t) \in \mathcal{R}^{m \times n}$ is a full row-rank matrix,
and $c(t)\in \mathcal{R}^{n}$ and $b(t)\in \mathcal{R}^{m}$
are the given vectors.

The solution of the minimization problem \re{fx1} changes with time,
and we expect to find it for each time. As such,
we have to consider this problem for each fixed instant $t' \in [0,\infty)$.
According to the conventional optimization theory,
we need to define the Lagrange function
$L(x(t'),\lambda(t'),t') =\frac{1}{2}x^{T}(t') G(t') x(t') +c^{T}(t')x(t')
+ \lambda^{T}(t')\left(A(t') x(t') - b(t')\right)
$
with $\lambda(t') \in R^{m}$ being the Lagrange-multiplier vector.
It is well known that
$(x^{\ast}(t'), \lambda^{\ast}(t') )$
is the optimal solution of \re{fx1},
if
$ \frac{\partial L(x, \lambda, t')}{\partial x}|_{x=x^{\ast}(t'),\lambda=\lambda^{\ast}(t')} = 0$
and $
\frac{\partial L(x,\lambda,t')}{\partial \lambda }|_{x=x^{\ast}(t'), \lambda=\lambda^{\ast}(t')} = 0$.
Going through the entire time interval,
the optimality condition for the time-variant QP problem \re{fx1}
can be rewritten in the matrix form, in view of the definition of $L$,
\bea \label{pro}
\mathbf{f}(z(t), t)  := u(t)-M(t)z(t)= 0
\eea
with
\bea {M(t)= \left[ {\begin{array}{*{15}{c}}
G(t)&A^{T}(t)\\
A(t)&\mathbf{0}_{m\times m}
\end{array}} \right]\in \mathcal{R}^{(n+m)\times(n+m)}} \nn
\eea
\bea {z(t) = \left[ {\begin{array}{*{15}{c}}
x(t)\\
\lambda(t)
\end{array}} \right]\in \mathcal{R}^{n+m},
u(t)= \left[ {\begin{array}{*{15}{c}}
-c(t)\\
b(t)
\end{array}} \right]\in \mathcal{R}^{n+m}} \nn
\eea
where $z(t)$ is the unknown vector to be solved,
$u(t)$ represents the given vectors,
and $M(t)$ is the known coefficient matrix.
Since $G(t)$ is positive definite and $A(t)$ is full row-rank,
then $M(t)$ is invertible. Eq. \re{pro} is in turn consistent
and a unique solution exists for each time instant.
In fact, the theoretical solution of Eq.\re{pro} can be given as $z^{\ast}(t) = M^{-1}(t) u(t)$,
where the inverse of $M(t)$ is required
to be available.
In particular, the theoretical solution
$z^{\ast}(t) = [x^*(t), \lambda^{*}(t)]$ with
$x^*(t) = -G^{-1}(t)(c(t)+A^{T}(t)\lambda^{*}(t))$
and $\lambda^{*}(t)
= -(A(t)G^{-1}(t)A^{T}(t))^{-1}(A(t)G^{-1}(t)c(t) +b(t))$.

In order to avoid the matrix inversion,
the ZNN approach
is suggested to solve the time-variant problem \re{pro}, in the absence of $\mathbf{w}$.
The rectifying action is designed
to make the error variable be enforced to zero,
by which the solution can be obtained.
Let us introduce the error $\mathbf{e}:= Mz-u$.
The computing objective is
to find the vector-valued function $z$
through zeroing the error $\mathbf{e}$.
The parameter setting errors, calculating  errors (adding, subtracting, multiplying and calculating derivatives)
and external disturbances
are inevitable in the implementation.
With ERS \re{err.dyn},
the disturbances occurred in practice can be coped with in a lumped way
and
the solving process of Eq. \re{pro}
can be characterized and monitored.
As such, the resultant ZNN model is definitely uncertain, described by
\bea \label{model}
M(t)\dot{z}(t)&=&-\dot{M}(t)z(t)+\dot{u}(t)
+\mathbf{r}\left(u(t)-M(t)z(t)\right) \ne
&&+\mathbf{s}\left(u(t)-M(t)z(t)\right) + \mathbf{w}(t)
\eea

ZNN provides an innovative alternative approach
 that has been widely applied for solving various types of time-variant zeroing problems \cite{zhang16}.
The methodology is `zeroing', i.e., making each element of a vector/matrix-valued error function
to vanish
by adopting a pre-specified error dynamics.
The error function can be defined, according to the computing problem to be solved,
and
it is governed by the error dynamics.
Defining the error function leads to
an implicit dynamics of
the ZNN model undertaken.
Different from the explicit dynamics by gradient descent method, this approach assures that
the computing error converges to zero,
obeying the evolution of the error dynamics,
so that the solution can be achieved.
The error dynamics in fact plays a crucial role
to assure stability, convergence
and robustness
of the computing scheme in a global sense,
and the major task is
the performance analysis based on the given error dynamics.
It has long been established that various ZNN designs, in the absence/presence of disturbances, are available from the published literature.
Most of the reported works show effectiveness of the direct way to construct ideal error dynamics,
with disturbance rejection techniques including
the integral action, the discontinuous action, etc.
Inspired by
the success of exploiting ZNN computing schemes,
systematic and constructive
design methodologies in control-theoretical perspective would be desirable and helpful,
to meet the demand of solving
complicated scientific computation problems.
In \cite{jin17}, the effort has been made
under the framework presented in \cite{bhaya07}.
It was shown the integral action is capable of attenuating bounded disturbances.

The benefit from ERS approach is the constructive design, in order to cope with disturbances,
and computing schemes can be realized by the resultant ZNNs.
The rectifying action and the robustifying term are specially involved in the error equation, in order to improve the convergence rate and enhance robustness with respect to varieties form of uncertainties.
To meet the requirement for the time-variant computing,
it is practically important
to shorten the settling time
or give a precise estimation for it.
In this paper,
novel rectifying actions are formed
so that the fixed-time settling can be achieved,
by which both the convergence rate
and the computing accuracy can be further improved.
Moreover, from a control-theoretical viewpoint,
signum/smoothing-signum actions are adopted
for robustness improvement.
In comparison with the existing works,
the convergence performance of
our computing schemes
can be improved dramatically,
according to the
derived settling time functions,
and their
robustness can be improved further,
under the deliberately introduced framework.

We give the following technique lemma
of finite-duration stabilization,
being helpful for design and analysis of  robust ZNN models to be presented.

\begin{lemma}
\label{lem.RAtool}
For the positive definite function $V$ satisfying that
\bea
\dot V \leq - (K+R) V^\alpha + \Delta,~~V(0)=V_0
\eea
with the given $\alpha \ge 0$ and $\Delta >0$,
and $K > 0$ and $R>0$ being the adjustable gains,
 then
$V$ converges to and remains within the region of the origin, bounded by $\left (\frac{\Delta}{R} \right )^{1/\alpha}$,
for which the needed time is at least,

i) for $\alpha < 1$,
$t_\Delta = \frac{1}{K(1-\alpha)} \left (
 V_0^{1-\alpha}-
\left (\frac{ \Delta}{ R} \right )^{1 -\alpha}
\right )$;

ii) for $\alpha = 1$,
$t_\Delta = \frac{1}{K} {\rm ln}
\left ( \frac{R}{\Delta} V_0  \right )$; and

iii) for $\alpha > 1$,
$t_\Delta = \frac{1}{K(\alpha -1)} \left ( \left (\frac{R}{\Delta} \right )^{\alpha -1} - \left (\frac{1}{V_0} \right )^{\alpha -1}  \right )$.

\end{lemma}

Lemma \ref{lem.RAtool} plays a key role in system designs,
which represents application of
the concept of practical stability. 
It should be noted that for the purpose of stabilization,
the restriction of $0 < K, R < 1$ is relaxed.

\section{ERS design using power laws } \label{sec.pral}

In this section,
for ERS designs,
let us first consider
ZNN model \re{model} in the absence
of disturbance, i.e., $\mathbf{w}=0$ (in turn $\mathbf{s}=0$).
The rectifying action $\mathbf{r}$ is constructed, in order to assure convergence performance of the ZNN model undertaken.
We would like to carry out a redesign procedure in Section \ref{sec.rres},
for choosing $\mathbf{s}$
in the presence of $\mathbf{w}$.

\subsection{Exemplary finite-time stable systems as attracting laws}

The dynamic behaviour of a ZNN model
usually obeys an attracting law.
By an attracting law we mean an exemplary dynamical system,
under which the dynamics of the ZNN is governed.
In this paper,
exemplary finite-time stable systems
that are used for ZNN model designs
are suggested as attracting laws,
for example, power laws.
Behaviour of these models is of
global terminal attractivity,
and in this paper
we call them terminal zeroing neural networks (TZNNs).
For details about the concepts of finite/fixed-time stability,
we refer to literature \cite{polyakov12,efimov21} and references therein.

Consider the following nonlinear system:
\[
\dot x = f(x), ~~x(0)=x_0
\]
with $f(0) =0$.
The origin is asymptotically stable,
if the right-hand side function $f$ is
monotonically increasing and odd;
the origin is finite-time stable,
if the settling time function satisfies
\[
t_s(x_0) = -\int_0^{x_0} \frac{d x}{ f(x)} < \infty\]
furthermore, the origin is fixed-time stable, if the settling time function
is bounded with respect to the initial value $x_0$,
satisfying that
\[\sup_{x_0} t_s(x_0) < \infty
\]

For each computing scheme, we construct the specific rectifying action,
on the basis of the existing and recently developed power laws, which are
listed in Tab \ref{tab.af}.
We provide
TZNN designs associated with settling time functions and more close estimates of settling time of these attracting laws.

\begin{table*}[!htbp]
\caption{Power laws}
\centering
\begin{tabular}{llll}
\hline
\\
\multicolumn{1}{p{.2in}}{AL}
&\multicolumn{1}{p{1.0in}}{$f(x)$}
&\multicolumn{1}{p{1.1in}}{Parameters}
&\multicolumn{1}{p{0.7in}}{Terminal value}\\
\hline
\\
SPRL
& $
- \kappa {\rm sig}^{\gamma}(x)
$ &
$ \kappa >0, 0<\gamma<1
$  &finite-time
\\
${\rm SPRL_{a.l.t.} }$
& $- \rho x - \kappa {\rm sig}^{\gamma}(x)$
& $ \rho > 0$ & finite-time
\\
DPRL
&
$-\kappa_1 {\rm sig}^{\gamma_1}(x) - \kappa_2 {\rm sig}^{\gamma_2}(x)$
& $\kappa_1, \kappa_2 >0, 0<\gamma_1<1, \gamma_2 >1$ & fixed-time
\\
${\rm DPRL_{a.l.t.} }$
&
$- \rho x -\kappa_1 {\rm sig}^{\gamma_1}(x) - \kappa_2 {\rm sig}^{\gamma_2}(x)$
& $\rho > 0$ & fixed-time
\\
TPAL
& $
- \kappa {\rm sig}^{\gamma(x) }(x), \gamma(x) =
\left\{ \begin{array}{ll}
\gamma_1,
&|x| < 1 \\
 \gamma_2 & |x| \ge 1
\end{array}
\right. $
&$\kappa >0, 0<\gamma_1<1, \gamma_2 >1$
& fixed-time
\\
${\rm TPAL_{a.l.t.} }$
& $
- \rho x - \kappa {\rm sig}^{\gamma(x) }(x) $
&$\rho>0$
& fixed-time
\\
\hline
\end{tabular}
\label{tab.af}
\end{table*}

\subsection{Single power-rate laws}

To illustrate the TZNN design,
we would like to begin with
one simple attracting law, a single power-rate law,
for simplicity of the presentation
and understandability.
 Let us choose the following form of rectifying action:
\bea
\label{sprr1}
 r_i(e_i) = - \kappa {\rm sig}^{\gamma}(e_i)
\eea
where $\kappa > 0$
and $0 < \gamma <1$;
$r_i, i=1,2,\cdots,n$, is the $i$th element of the rectifying action ${\bm r}$.
The equilibrium of the error dynamics with the use of \re{sprr1} is the origin, $e_i = 0$.
The settling time of
the TZNN can be derived as
\bea
\label{sprr1.ts}
 t_s(e_0) = \frac{1}{\kappa(1-\gamma)}
 |e_0|^{1-\gamma}
\eea
Special choices of $\gamma$ are needed  for the particular implementation.
Let us set $\kappa = 1$, and
for $\gamma = \frac{1}{2}$,
\bea
\label{sprr1.d5}
 t_s(e_0) = 2
 \sqrt{|e_0|}
\eea
for $\gamma = \frac{1}{3}$,
\bea
\label{sprr1.d3}
 t_s(e_0) = \frac{3}{2}
 |e_0|^{\frac{2}{3}}
\eea
and for $\gamma = \frac{1}{n},~ n >1$,
\bea
\label{sprr1.dn}
 t_s(e_0) = \frac{n}{n-1}
 |e_0|^{\frac{n-1}{n}}
\eea
Settling function \re{sprr1.dn} becomes $|e_0|$, as $n$ increases.

It is well-known that by adding a linear term ( a. l. t., for short), the convergence rate can be
potentially sped up \cite{yu18}.
Thereby, we apply the two-term attracting law by choosing the rectifying action
\bea
\label{sprr2}
r_i(e_i) = - \rho e_i - \kappa
{\rm sig}^{\gamma}(e_i)
\eea
where $\rho >0$, $\kappa > 0$,
and $0 < \gamma <1$.
The settling time function of
this TZNN can be given by
\bea
\label{sprr2.ts}
t_s(e_0) = \frac{1}{\rho(1-\gamma)}
{\rm ln} \left ( \frac{\rho}{\kappa} |e_0|^{1-\gamma} +1 \right )
\eea
Note that ${\rm ln} (1+x) \le x,$ for $ x> -1$ and $x \neq 0$.
By comparing \re{sprr1.ts} and \re{sprr2.ts},
a faster convergence rate by \re{sprr2}
is achieved than that by \re{sprr1}, and it is observed that
the technique of adding a linear term
can really speed up the convergence rate.
Unfortunately,
it also follows from \re{sprr1.ts} and \re{sprr2.ts}
that the settling time will increases with respect to $|e_0|$.

\subsection{Double power-rate laws}

One efficient design for $r_i(e_i)$ can be conducted with the use of double power-rate laws.
The following  properties of special functions are helpful for the settling time derivation and estimation to be presented:
$B(x,1-x) = \Gamma(x)\Gamma(1-x) = \frac{\pi}{\mathrm{sin}(\pi x)} $ and $
I(x,p,q)=1-I(1-x,q,p)$,
where the incomplete Beta-function $I(x,p,q)$ is defined by the integral, $I(x,p,q) = {1}/{B(p,q)}\int_0^{x}t^{p-1}(1-t)^{q-1}dt $, for $0 \leq x < 1$, and
$B(p,q) = \int_0^{1}t^{p-1}(1-t)^{q-1} dt$, for $p>0$ and $q>0$.

Let us choose the following form
of rectifying action:
\bea \label{qp}
r_i(e_i) = -\kappa_{1}{\rm sig}^{\gamma_1}(e_i) -
\kappa_{2}{\rm sig}^{\gamma_2}(e_i)
\eea
where $\kappa_1 > 0$, $\kappa_2>0$,
$0 < \gamma_1 <1$, and
$\gamma_2 >0$.
The equilibrium of the error dynamics with the use of \re{qp} is the origin $e_i = 0$.

\begin{theorem} \cite{sun23}
\label{thm.qp}
Given an initial error $e_0$,
in the absence of $\mathbf{w}$,
TZNN model \re{model} together with \re{qp}
is finite-time table,
associated with the following settling time function
\bea
t_s(e_0) =
\frac{\pi \mathrm{csc}(\theta\pi)}{\kappa_{1}(\gamma_2-\gamma_1)}
\left(\frac{\kappa_{1}}{\kappa_{2}} \right)^{\theta}
I(1-c,\theta,1-\theta)
\label{key.ts}
\eea
where $\theta = (1-\gamma_1)/(\gamma_2-\gamma_1)$ and
$c=\kappa_{1}/(\kappa_{2} |e_0|^{\gamma_2-\gamma_1}+\kappa_{1})$.
\end{theorem}

The expression in Theorem \ref{thm.qp} is given with the aid of the incomplete beta function.
However,
it is not a
closed-form one and
difficult to find the exact value because of the difficulty in computing the special function, especially for realizing a predefined-time stable RNN design.
It makes sense that
the finite-time settling exists
for each given initial condition.
By choosing $\gamma_1 + \gamma_2 = 2$, the closed-form expression can be obtained as  \cite{zuo14}
\bea
\label{key.ts.2}
t_s(e_0)
= \frac{1} { \sqrt{\kappa_1 \kappa_2}(1-\gamma_1)} \arctan \left ( \sqrt{ \frac{\kappa_2}{\kappa_1} }  |e_0|^{1-\gamma_1}\right )
\eea
and for $ 2\gamma_1 + \gamma_2 = 3$,
\bea
\label{key.ts.3}
t_s(e_0)
&=& \frac{1} { 6 \kappa_2 a^2(1-\gamma_1)} \bigg (
{\rm ln} \frac{|e_0|^{2(1-\gamma_1)} + 2 a |e_0|^{1-\gamma_1} +a^2}
{|e_0|^{2(1-\gamma_1)} - a |e_0|^{1-\gamma_1} +a^2   }
 \ne && + 2 \sqrt{3}\arctan
 \left (
\frac{2 \sqrt{3}}{3a}
   \left (|e_0|^{1-\gamma_1} -\frac{a}{2} \right )\right )
   + \frac{\sqrt{3}\pi}{3}
\bigg )
\eea
with $a = \left ( \kappa_1 /\kappa_2 \right )^{\frac{1}{3}}$.

Nevertheless, an estimate of uniform bound on the settling time function with respect to initial condition
was given as \cite{hu22}:
\bea
\label{key.ts.bound}
t_s(e_0) \leq \frac{\pi \rm{csc}(\theta \pi)}{\kappa_1(\gamma_2-\gamma_1)} \left(\frac{\kappa_1}{\kappa_2} \right)^{\theta}
\eea
by which TZNN model \re{model} with \re{qp} is assured to be fixed-time stable.
The expressions for typical choice of the power rates can be derived.
By choosing $\gamma_2 - \gamma_1 = n (1-\gamma_1)$,
\bea
\label{key.ts.bound.n}
t_s(e_0) \leq \frac{\pi } {n \kappa_1 (1-\gamma_1) {\rm sin}(\pi/n) } \left(\frac{\kappa_1}{\kappa_2} \right)^{\frac{1}{n} }
\eea
especially, for $n =2 $ \cite{zuo14},
\bea
\label{key.ts.bound.2}
t_s(e_0) \leq \frac{\pi} {2 \sqrt{\kappa_1 \kappa_2}(1-\gamma_1) }
\eea
and
for $n =3 $,
\bea
\label{key.ts.bound.3}
t_s(e_0) \leq \frac{2 \sqrt{3} \pi} {9 \kappa_2 (1-\gamma_1) }
\left (
\frac{\kappa_1}{\kappa_2} \right )^{- \frac{2}{3}}
\eea

\subsection{Double power-rate laws of adding a linear term}

By adding the linear term $\rho x, \rho >0,$ in \re{qp},
a convergence-rate-enhanced TZNN can be obtained by applying
\bea \label{rqp}
r_i(e_i)= - \rho e_i -
\kappa_{1}{\rm sig}^{\gamma_1}(e_i) -
\kappa_{2}{\rm sig}^{\gamma_2}(e_i)
\eea

The linear part plays an essential role in the improvement of convergence rate of the error system.
When applying \re{rqp},
it is difficult to directly calculate
the improper integral given by  $\int_0^{+\infty}1/(\rho e_i +\kappa_{1} e_i^{\gamma_{1}}
+\kappa_{2} e_i^{\gamma_{2}})de_i$.
With the aid of the special function,
a novel estimation of settling time of
TZNN model \re{model}-\re{rqp} can be conducted.

\begin{theorem}
\label{thm.rqp.ts}
The settling time of
TZNN model \re{model} together with \re{rqp}  is
estimated as
\bea
\label{rqp.tg}
t_s(e_0)\!\!\!&\leq& \!\!\!
 \frac{\pi \mathrm{csc}(\theta\pi)}{\kappa_1(\gamma_2-\gamma_1)}
\left(\frac{\kappa_1}{\rho+\kappa_{2}} \right)^{\theta} I(1-c,\theta,1-\theta)  + \frac{\pi \mathrm{csc}(\theta\pi)}
{\kappa_2(\gamma_2-\gamma_1)}\left( \frac{\kappa_2}{\rho+\kappa_{1}}\right)^{1-\theta} I(b,1-\theta,\theta)
\eea
where
$\theta=(1-\gamma_1)/(\gamma_2-\gamma_1)$,
$b=(\rho+\kappa_1)/(\rho+\kappa_1+\kappa_2)$,
and $c=\kappa_1/(\rho+\kappa_1+\kappa_2)$.
\end{theorem}

{\bf Proof.}
Let us choose the Lyapunov function candidate $V(e_i) = |e_i|$.
The settling time
can be estimated in two phases.
For the traveling phase (from $e_{i}(0) \ge 1$
to $e_{i}(t_1) = 1$),
we have
$\dot{V}_{i}\leq - (\rho+\kappa_1 )V_i^{\gamma_1}
 -\kappa_2 V_{i} ^{\gamma_2}$.
It follows that
\bea
t_1\leq \int_1^{+\infty}\frac{1}{(\rho+\kappa_1) V_{i}^{\gamma_1}
+\kappa_2 V_{i} ^{\gamma_2}}d V_i  \nn
\eea
Let us define
$ u = \frac{\rho+\kappa_{1}} {\kappa_2 V_i^{\gamma_2-\gamma_1} + \rho+\kappa_{1}}$,
which leads to
$e_i=
\left(\frac{\rho+\kappa_{1}}{\kappa_2}\right)^{\frac{1} {\gamma_2-\gamma_1}}
\left(\frac{1}{u} - 1\right)^{\frac{1} {\gamma_2-\gamma_1} }$
and
$ \frac{1}{(\rho+\kappa_{1})V_i^{\gamma_1}
+\kappa_2 V_i^{\gamma_2}}
=\frac{1}{\rho+\kappa_{1}} u V_i^{-\gamma_1}$.
Then we obtain
\bea    \label{max.t1}
t_1
\!\!\!\!\!&\leq&\!\!\!\!\! \frac{1}
{(\rho+\kappa_{1})(\gamma_2-\gamma_1)}\left(\frac{\rho+\kappa_{1}} {\kappa_2}\right)^{\theta}
\int_{0}^{b} u^{-\theta}(1-u)^{\theta-1} du \ne
\!\!\!\!&=&\!\!\!\!\!\!\! \frac{1}
{\kappa_2(\gamma_2-\gamma_1)}\left(\frac{\kappa_2}{\rho+\kappa_{1}}\right)^{1-\theta}
B(1-\theta,\theta)                       I(b,1-\theta,\theta)
\eea

Similarly, for the arrival phase (from
$e_{i}(t_1)= 1$ to $e_{i}(t_2)=0$),
we have
$ \dot{V}_{i}\leq- \kappa_1 V_i^{\gamma_1}-(\rho+\kappa_2) V_i^{\gamma_2}$.
Defining
$w = \frac{\kappa_1} {(\rho+\kappa_2) V_i^{\gamma_2-\gamma_1}+ \kappa_1}$ gives rise to
\bea  \label{max.t2}
t_2 \!\!&\leq&\!\! \int_{0}^{1}\frac{1}{\kappa_1 V_i^{\gamma_1}
+(\rho+\kappa_2) V_i^{\gamma_2}}d V_i  \ne
\!\!&=&\!\! \frac{1}{\kappa_1(\gamma_2-\gamma_1)}
\left(\frac{\kappa_1}{\rho+\kappa_{2}}\right)^{\theta}
B(1-\theta,\theta)                           \left(1-I(c,1-\theta,\theta)\right)
\eea
Hence, boundedness of the settling-time function follows by
combining \re{max.t1} and \re{max.t2}.
\QED

The bound estimation was found to be overestimated
 in many existing works,
where the bound estimation of the settling time of \re{rqp}
is replaced by that of \re{qp}.
In Theorem \ref{thm.rqp.ts},
we take into account the influence of power- and linear-terms in different
phases, and establish a more accurate
estimate for the upper bound.

For the case of $\gamma_{2}+\gamma_{1}=2$ in \re{rqp}, the perfect estimation for settling time can be made by the closed-form expression of $t_s(e_0)$.

\begin{theorem}
\label{thm.rqp}
For the parameter setting $\gamma_{2}-\gamma_{1}=2(1-\gamma_{1})$,
the
TZNN model \re{model} with \re{rqp} is finite-time stable,
and the settling time function can be given
in terms of
$a=4\kappa_1\kappa_2-\rho^2$.
For $a>0$,
\bea    \label{1qp.ts.1}
t_s(e_0)\!\!&=&\!\! \frac{2}{(1-\gamma_1)\sqrt{a}}
 \arctan \left ( \frac{\sqrt{a}|e_0|^{1-\gamma_1} }{2\kappa_1 + \rho|e_0|^{1-\gamma_1}} \right )
\eea
satisfying that
\bea
\label{1qp.tc.1}
t_s(e_0) \le \frac{2}{(1-\gamma_{1})\sqrt{a}} \arctan\left( \frac{\sqrt{a}}{\rho}  \right);
\eea
for $a<0$,
\bea    \label{1qp.ts.2}
t_s(e_0)
= \frac{1}{(1-\gamma_{1})\sqrt{-a}}
{\rm ln}\left(\frac{\left ( \rho+\sqrt{-a} \right )
\left ( 2 \kappa_2 |e_0|^{1-\gamma_{1}}+ \rho-\sqrt{-a} \right ) }{\left ( \rho-\sqrt{-a} \right )
\left ( 2 \kappa_2 |e_0|^{1-\gamma_{1}}+ \rho+\sqrt{-a}\right ) }\right)
\eea
satisfying that
\bea   \label{1qp.tc.2}
t_s(e_0) \le \frac{1}{(1-\gamma_{1})\sqrt{-a}}
{\rm ln}\left(\frac{\rho+\sqrt{-a}}
 {\rho-\sqrt{-a}}\right);
\eea
and for $a=0$,
\bea \label{1qp.ts.3}
t_s(e_0) = \frac{4}{(1-\gamma_1)\rho} \frac{\kappa_2 |e_0|^{1-\gamma_1 }}{  2 \kappa_2 |e_0|^{1-\gamma_1 }+\rho }
\eea
satisfying that
\bea \label{1qp.tc.3}
t_s(e_0) \le \frac{2}{\rho(1-\gamma_{1})}.
\eea
\end{theorem}

The rectifying action \re{rqp} becomes \re{qp}, as $\rho=0$ ($a>0$).
It is seen that the convergence rate of the TZNN undertaken is improved due to the introduction of a linear term in \re{qp}.
It is seen that when $\rho=0$, the settling time function and upper bound given in
\re{key.ts.2}
and \re{key.ts.bound.2}
 can be obtained by \re{1qp.ts.1} and \re{1qp.tc.1}, respectively.

Our study about the three-term attracting law \re{rqp} was conducted, especially for the closed-form expression representing its settling time function,
in terms of
$a=4\kappa_1\kappa_2-\rho^2$,
and the derivation
for the case $a>0$
and the related stability result
was reported in \cite{wang21}.
It should be noted that
the case study by assuming $a=0$ was found in \cite{zhangzeng22},
which causes many concerns recently
about various finite-time stability problems.
Here we present a similar result.
To the best of our knowledge,
in Theorem \ref{thm.rqp},
the expression and estimation
of the settling time in the situation where $a<0$
is reported for the first time.

\section{ERS design using power-exponential laws} \label{sec.tpAL}

An alternative way, in this section,
is provided for fixed-time stabilization of the constructed TZNNs.
We introduce the class of power-exponential attracting laws,
described by the following differential equation:
\[
\dot x = -\kappa f^{g(x)}(x){\rm sgn}(x)
\]
where $\kappa>0$,
$f(x)$ is the power base,
$g(x)$ is the power exponent, and
both of these are allowed to be functions of $x$.
We assume that the origin is an equilibrium solution.
In fact,
various power-exponential laws exist
for the time-varying computing purpose,
by choosing appropriate functions $f(x)$ and $g(x)$.
The benefit from the introduction of power-exponential laws is that we can obtain the closed-form of settling-time function of the designed TZNN
or derive the exact estimation of settling time.
The adding-linear-term technique is applicable for performance improvement, by which the system dynamics undertaken can be given as
\[
\dot x = -\rho x -\kappa f^{g(x)}(x){\rm sgn}(x)
\]
with $ \rho>0$.
With such power-exponential laws.
we will show that the logarithmic settling-time
can be achieved.

Two-phase fixed-time convergent systems are presented, in \cite{sun20}, having one power-rate term and the exponent takes two values that are
greater than or less than 1, determined by the transition state.
It was clarified in \cite{sun21b} that
the convergence rate is possible to be sped up,
if the transition state is not set to 1.
In this paper,
we will continue to study and explore
novel form of such TZNNs,
by which the fixed-time convergence results are
particularly beneficial to the time-variant problem solving.

We consider the typical power-exponential rectifying action expressed by
\bea
r_i(e_i) &=& - \rho \left(\frac{e_i}{e^{*}}\right) - \kappa \left(\frac{|e_i|}{e^*}\right)^{\gamma(e_i)}{\rm sgn}(e_i), \label{sys.rhope}\\
\gamma(e_i) &=& \left\{
\begin{array}{lll}
\gamma_{1}, & \mbox{$ |e_i| < e^*$},\\
\gamma_{2}, & \mbox{$ |e_i| \ge e^*$},
 \end{array}
\right.
\label{fda.qp.gamma-2}
\eea
 where $\rho \geq 0$, $\kappa > 0$,
 $0<\gamma_{1}<1$,
$\gamma_{2}>1$,
and $e^*>0$ is the transition state.
The Heaviside-like function
for constructing $\gamma(e_i)$ is piecewise discontinuous.
However, the right-hand side function of power-exponential rectifying action \re{sys.rhope}
with the specified exponent is continuous at the transition state $e^*$,
and the continuity assures the existence and
uniqueness of the solution of the corresponding TZNN.

\begin{theorem}
\label{thm.gammaC}
Consider
TZNN model \re{model} together with
\re{sys.rhope} and \re{fda.qp.gamma-2}.
The expression of settling time function can be derived as

i) for $\rho>0$, $|e_0|>e^{*}$
\bea
t_{s}(e_0)
=  \frac{e^{*}}{\rho (1-\gamma_1)}{\rm ln} \left ( 1+\frac{\rho}{\kappa}  \right )
 + \frac{e^{*}}{\rho (\gamma_2-1)}  {\rm ln} \left( \frac{1+\frac{\kappa}{\rho}}
{\left(\frac{|e_0|}{e^{*}}\right)^{1-\gamma_2}+\frac{\kappa}{\rho} }\right)     \nn
\eea
as $0 < \vert e_0 \vert < e^{\ast}$, then
\bea
t_{s}(e_0) =
  \frac{e^\ast}{\rho(1-\gamma_1)}
{\rm ln}\left ( 1+\frac{\rho}{\kappa} \left(\frac{\vert e_0\vert }{e^{\ast}}\right)^{1-\gamma_1} \right )  \nn
\eea
ii) for  $\rho=0$, $|e_0|\geq e^*$
\bea
t_s(e_0)=\frac{e^*}{\kappa(1-\gamma_{1})}+\frac{e^*}{\kappa(\gamma_{2}-1)}-\frac{(e^*)^{\gamma_{2}}}{\kappa(\gamma_{2}-1)}|e_0|^{1-\gamma_{2}}  \nn
\eea
as $|e_0|< e^*$, then
\bea
t_s(e_0)=\frac{e^*}{\kappa(1-\gamma_{1})}\left(\frac{|e_0|}{e^*}\right)^{1-\gamma_{1}}  \nn
\eea
\end{theorem}
{\bf Proof.} The proof is similar to that for Lemma 2 in \cite{sun20} and omitted
here due to space limitation. \QED

It should be noted that the closed-form expression of settling time function is obtained for each given initial condition,
and with this expression,
the settling time can be exactly computable.
In comparison to \re{sys.rhope} with $\rho =0$,
the convergence rate is improved, due to
the logarithmic settling time
realized by adding the linear term
\re{fda.qp.gamma-2}.
In Theorem \ref{thm.gammaC},
we obtain the settling time function
with respect to the specified transition state,
whereas in \cite{sun20},
the transition state is set to $1$.
The settling time underlines the selection of the transition state.
The transition state $e^*$ of \re{sys.rhope} provides an adjustable factor, and
the settling time can be decreased
by reducing $e^*$.
More importantly,
by choosing appropriate parameters, the convergence rate when using the two-phase rectifying action \re{sys.rhope} can be faster than that when using \re{rqp}.
The comparison result is presented in the following corollary.

\begin{corollary}
\label{compare}
By setting $e^*=1$,
and choosing $\kappa = \kappa_1+\kappa_2$ in \re{sys.rhope},
the settling time by using \re{sys.rhope}
 is shorter than that by applying \re{rqp}.
\end{corollary}

{\bf Proof.}
It follows from the error dynamics with \re{sys.rhope} that for $e_{0}\geq 1$,
$\dot{e}_{i} \leq - \rho e_{i} - \kappa_1 e_{i}^{\gamma_2}- \kappa_2 e_{i}^{\gamma_2}$;
and for $0<e_{0}<1$,
$\dot{e}_{i} \le - \rho e_{i} - \kappa_1 e_{i}^{\gamma_1}-\kappa_2 e_{i}^{\gamma_1}$.
As for \re{rqp},
for $e_{0}\geq 1$,
 we have,
$e_{i}^{\gamma_1}\leq e_{i} \leq e_{i}^{\gamma_2}$. Then
$\dot{e}_{i}\geq  - \rho e_{i} - \kappa_1 e_{i}^{\gamma_2} -\kappa_2 e_{i} ^{\gamma_2}$.
For $0<e_{0}<1$, we have $e_{i}^{\gamma_2}\leq e_{i}\leq e_{i}^{\gamma_1}$. Then
$
\dot{e}_{i}\geq  - \rho e_{i} - \kappa_1 e_{i} ^{\gamma_1} - \kappa_2 e_{i}^{\gamma_1}$.
Hence, the corollary follows according to  the above inequalities for \re{rqp} and
\re{sys.rhope}.
\QED

The $\gamma(e_i)$ of the two-phase exponent \re{fda.qp.gamma-2} is discontinuous at $e_i = e^*$,
resulting in the non-smooth (but continuous) right-hand side function.
Such discontinuous exponent would determine attractive performance of the TZNN undertaken, which deserves further exploration.
It may lead to
difficulty in the implementation,
because certain applications in real-time computing problems require the right-hand
side function to be continuous or smooth.
We shall illustrate that
the exponent given by \re{fda.qp.gamma-2} is not exclusive
and there are alternatives for constructing rectifying actions.
Let us consider that $\gamma(e_i)$ is chosen as a continuous function to assure the smoothness of the right-hand side function.
The idea is to design the continuous $\gamma(e_i)$ that
approximates the exponent \re{fda.qp.gamma-2}.
We shall show that
fixed-time stability of the constructed TZNN models is assured with the use of the state-dependent exponents.

Among others,
two power-exponential rectifying actions are those
having the following exemplar exponents, with the transition state $e^*$  to be specified by designer,
\bea
\gamma(e_i) &=& \left\{
\begin{array}{lll}
\gamma_{1}, & \mbox{$ |e_i| \leq \delta e^*$},\\
\gamma_{01}(e_i), &\mbox{$ \delta e^* < |e_i| < e^*$},\\
\gamma_{2}, & \mbox{$ |e_i| \ge e^*$},
 \end{array}\label{fda.qp.gamma}
\right.
\eea
and
\bea
\gamma(e_i) &=& \left\{
\begin{array}{lll}
\gamma_{1}, & \mbox{$ |e_i| \leq e^*$},\\
\gamma_{02}(e_i), &\mbox{$ e^*< |e_i| < e^* +\delta e^* $},\\
\gamma_{2}, & \mbox{$ |e_i| \ge e^*+\delta e^* $},
 \end{array}  \label{fda.qp.gamma-1}
\right.
 \eea
where $0<\delta<1$, $e^*>0$,
$\gamma_{01}(e_i)=(\gamma_{2} - \gamma_1)/(1-\delta)(e_i/e^*-1)+\gamma_{2}$,
and
$\gamma_{02}(e_i)=((\gamma_{2}-\gamma_{1})/\delta )(e_i/e^*-1)+\gamma_{1}$.
The exponents are continuous, assuring that the derivative of right-hand side function of \re{sys.rhope} is continuous.

With similar derivations to Theorem \ref{thm.gammaC},
the stability of the TZNN models with the given power-exponential rectifying action
is respectively established in the following theorem.
\begin{theorem}
\label{lem.fda.fastqp}
Applying the rectifying action \re{sys.rhope} together with the exponent \re{fda.qp.gamma},
the settling time function satisfies that
\bea
\label{fastqp.t1t2}
t_s(e_0)
\le \frac{e^*}{\rho (1-\gamma_1)}
{\rm ln} \left (1 + \frac{\rho}{\kappa} \delta^{1-\gamma_1}
   \right) + \frac{e^*}{\rho (\gamma_2-1)}
{\rm ln} \left (1 + \frac{\rho}{\kappa} \delta^{1-\gamma_2}
   \right)
\eea
And with the exponent \re{fda.qp.gamma-1},
 the settling time function satisfies that
\bea
\label{fastqp.t1t2.b}
t_s(e_0)
\le  \frac{e^*}{\rho (1-\gamma_1)}
{\rm ln} \left (1 + \frac{\rho}{\kappa} (\delta+1)^{1-\gamma_1}
   \right)    + \frac{e^*}{\rho (\gamma_2-1)}
{\rm ln} \left (1 + \frac{\rho}{\kappa} (\delta+1)^{1-\gamma_2}
   \right)
\eea
\end{theorem}
The rectifying action \re{sys.rhope} is adopted by adding a proportional term.
With $\rho>0$,
the settling time function of \re{sys.rhope} realizes logarithmic fixed-time settling, due to the impact of the proportional term.
It is seen that the convergence rate can be improved dramatically,
by choosing appropriate transition state $e^*$ of \re{sys.rhope}.
In the case of taking $\rho=0$ in  \re{sys.rhope},
with $\gamma(e_i)$ being the same as \re{fda.qp.gamma} and \re{fda.qp.gamma-1},
the estimation on the settling time can be respectively given.

More power-exponential TZNNs are those with the exponent being a smooth function
proximate to the two-phase constant-exponent.
The fixed-time convergence perofrmance of the formed rectifying actions with the power-exponent can be assured, because of the proximation to Heaviside-like function \re{fda.qp.gamma-2}.
Here we propose novel fractional rectifying action in the form of
\bea
\gamma(e_i) = \frac{\alpha+\beta \left(\frac{|e_i|}{e^{*}}\right)^{2m}}{1+ \left(\frac{|e_i|}{e^{*}}\right)^{2m}} \label{gamma}
 \eea
where $0\leq \alpha<1$, $\beta>2$,
and $m$ is a positive integer to be chosen.
On the one hand, \re{sys.rhope} with  \re{gamma} is equivalent to
$ r_i(e_i) = \rho\left(\frac{e_i}{e^*} \right)+ \kappa\left|\frac{e_i}{e^*}\right|^{\beta} {\rm sgn}(e_i)$,
as $e_i\rightarrow \infty$;
on the other hand,
$ r_i(e_i) =  \rho\left(\frac{e_i}{e^*} \right) +\kappa\left|\frac{e_i}{e^*}\right|^{\alpha} {\rm sgn}(e_i)$,
as $e_i \rightarrow 0$.

I

\begin{theorem}
\label{theorem.fastqfe}
For
the rectifying action \re{sys.rhope}
with the fractional exponent \re{gamma},
the settling
time function of TZNN model \re{model} satisfies that,
\bea
t_s(e_0) \le \frac{e^{*}}{\rho (\frac{\beta}{2}-1)}
{\rm ln} \left (1
  + \frac{\rho}{\kappa} \right )
 + \frac{e^{*}}{\rho (1-\alpha)}
{\rm ln} \left ( 1+\frac{\rho}{\kappa e^{\frac{-\beta}{2 m  {\rm e}}}}
 \right )
\label{b.fastqfe.t1t2}
\eea
And in the case of $\rho=0$,
the settling time function satisfies that,
\bea   \label{b.qfe.t1t2}
t_s(e_0)\leq \frac{e^{*}}{\kappa(\frac{\beta}{2}-1)}+\frac{e^{*}}{\kappa e^{\frac{-\beta}{2m  {\rm e}}}(1-\alpha)}
\eea
\end{theorem}

{\bf Proof.}
For $\rho>0$ in \re{sys.rhope},
the proof is presented by considering the traveling and reaching phases, respectively.

For the traveling phase (assuming that $|e_i| \ge  e^*$),
we have $\frac{\alpha+\beta \left(\frac{|e_i|}{e^{*}}\right)^{2m}}{1+ \left(\frac{|e_i|}{e^{*}}\right)^{2m}}\geq \frac{\beta \left(\frac{|e_i|}{e^{*}}\right)^{2m}}{1+ \left(\frac{|e_i|}{e^{*}}\right)^{2m}} \geq \frac{\beta}{2}>1$, leading to
$
\dot{e}_i \leq -\rho \left(\frac{e_i}{e^{*}}\right)
-\kappa\left(\frac{|e_i|}{e^{*}} \right)^{\frac{\beta}{2}}
$.
Defining $y = (|e_i|/e^{*})^{1-(\beta/2)}$ yields
$\dot y  \geq  - \frac{\rho}{e^{*}}\left ( 1-\frac{\beta}{2} \right )y   -\frac{\kappa}{e^{*}}  \left ( 1-\frac{\beta}{2} \right )$.
Solving it gives rise to
 \bea
 y(t)
  \!\!&\geq& \! \!e^{- \frac{\rho }{e^{*}} (1-\frac{\beta}{2}) t } y(0) - \frac{\kappa }{e^{*}}(1-\frac{\beta}{2})\int_0^t
 e^{- \frac{\rho }{e^{*}} (1-\frac{\beta}{2}) (t-s) } ds \ne
&\geq & \left ( y(0)
  + \frac{\kappa}{\rho} \right )
e^{- \frac{\rho }{e^{*}}(1-\frac{\beta}{2})t}  - \frac{\kappa}{\rho}
 \nn
\eea
Thus, for traveling from $y_0$ to $y(t_1) = 1$, the required time satisfies
\bea
 t_1 \leq \frac{e^{*}}{\rho  (\frac{\beta}{2}-1)}
{\rm ln} \left ( \left ( 1
  + \frac{\kappa}{\rho} \right ) / \left (y_0
  + \frac{\kappa}{\rho} \right ) \right)
\label{fastqfe.t1}
 \eea

For the reaching phase ($|e_i| \le  e^*$),
noting  $\frac{x}{1+x} \le 1$, for $x \ge 0$,
implying
$ \left(|e_i|/e^{*}\right)^{\gamma(e_i)}
\geq \left(|e_i|/e^{*}\right)^{\alpha+\beta \left(|e_i|/e^{*}\right)^{2m}}$.
Since ${\rm min}_{|e_i|} \left(\left(|e_i|/e^{*}\right)^{\beta \left(|e_i|/e^{*}\right)^{2m}} \right)
=e^{\frac{-\beta}{2m {\rm e}}}$,
then
$
\dot{e}_i \leq
-\rho \left(\frac{e_i}{e^{*}}\right)
-\kappa e^{\frac{-\beta}{2m {\rm e}}} \left(\frac{|e_i|}{e^{*}}\right)^{\alpha}
$.
Defining $y = (|e_i|/e^{*})^{1-\alpha}$ gives rise to
$ \dot y \leq -\frac{ \rho}{e^*} (1-\alpha)y - \frac{\kappa}{e^*}e^{\frac{-\beta}{2 m  {\rm e}}} (1-\alpha)$.
We obtain
\bea
 y(t)
&\leq& e^{- \frac{\rho}{e^*} (1-\alpha)t } y(t_1)
 - \frac{\frac{\kappa }{e^*} e^{\frac{-\beta}{2m e}}}{\rho} \left ( 1-
  e^{- \frac{\rho}{e^*} (1-\alpha)t } \right ) \ne
&=& \left ( y(t_1)
  + \frac{\kappa e^{\frac{-\beta}{2m  {\rm e}}}}{\rho} \right )
e^{- \frac{\rho }{e^*}(1-\alpha)t} - \frac{\kappa e^{\frac{-\beta}{2m  {\rm e} }}}{\rho}
 \nn
\eea
The time needed for
the phase from $y(t_1)=1$ to $y(t_2) =0$ is given as
\bea
t_2
  \leq \frac{ e^*}{\rho (1-\alpha)}
{\rm ln}  \left ( 1  + \frac{\rho}{\kappa e^{\frac{-\beta}{2 m  {\rm e}}}}
 \right )
\label{fastqfe.t2}
 \eea
Then \re{b.fastqfe.t1t2} can be obtained by
combining \re{fastqfe.t1} and \re{fastqfe.t2}.

As for $\rho=0$ in \re{sys.rhope},
the settling time is similarly derived through
the two-phase analysis.
For the traveling phase (assuming that $|e_i| \ge  e^*$),
$\dot{e}_i \leq -\kappa\left(\frac{|e_i|}{e^{*}} \right)^{\frac{\beta}{2}}$
and for $e_i(t_1)= e^*$,
\bea
 t_1 \le \frac{1}{\kappa} \frac{e^{*}}{\frac{\beta}{2} - 1}\left(1-\left(\frac{|e_0|}{e^{*}}\right)^{-\frac{\beta}{2}+1}\right)
\label{qfe.t1}
 \eea
For the reaching phase ($|e_i| < e^*$),
$
\dot{e}_i  \leq -\kappa e^{\frac{-\beta}{2m  {\rm e} }} \left(\frac{|e_i|}{e^{*}}\right)^{\alpha}$.
It needs for $e_i(t_2) = 0$,
\bea
t_2\leq  \frac{e^{*}}{\kappa e^{\frac{-\beta}{2 m  {\rm e}}} (1-\alpha)}\label{qfe.t2}
 \eea
Combining \re{qfe.t1} and \re{qfe.t2},
the estimate for settling time is $t_s(e_0) = t_1+t_2$, as given by \re{b.qfe.t1t2}.
\QED

The settling time functions for error dynamics with the fractional exponents
are bounded with respect to initial conditions, such that TZNN models with \re{sys.rhope}( for $\rho>0$ or $\rho=0$) are fixed-time stable.
Compared with case of $\rho=0$,
TZNN model in the case of $\rho \neq 0$ has
logarithmic settling-time estimation,
being able to be closer to the true value of the settling time function.
Note that the derivations are carried out in two-phase, i.e., $|e_i|>e^*$ and $0<|e_i|<e^*$.
The bound on the settling time function varies with the transition state $e^*$.
According to \re{b.fastqfe.t1t2} and
\re{b.qfe.t1t2}, to select an appropriate $e^*$ offers an speeding-up technique, by which the convergence rate can be boosted.
In addition,
$m$ can also effectively reduce the estimated  value of the bound of settling time.

\section{ERS redesign: Disturbance rejection and attenuation techniques}
\label{sec.rres}

%

In this section, the presence of uncertainty $
{\bf w} = \{w_i\}
$
 is taken into account,
by assuming that \[
|w_i|\leq \varpi
\]
with $ \varpi = \varpi(e,t) > 0$.
Under this assumption,
we introduce a compensation term in order
to reject the impact of $\bf{w}$ and thus
guarantee robust convergence of ERS,
while the rectifying action ${\bm r}$ is constructed in the former sections.
Now the remaining work is to construct the specific $\bf{s}$,
since we can apply the rectifying actions
in Section \ref{sec.res}.

The compensation term can be given through adopting the signum function:
\bea \label{s.sgn}
 s_i(e_i) = - \varpi \mathrm{sgn}(e_i)
\eea
where $s_i, i=1,2,\cdots,k$, is the $i$th element of $\bf{s}$.
Adopting the rectifying action \re{sys.rhope}
and the uncertainty compensation \re{s.sgn},
the $i$th ($i=1, 2, \ldots, k$ with $k=m+n$) neuron of
TZNN model \re{model} can be expressed as
\bea
\dot{z}_{i}
&=&\sum_{j=1}^{k} (I_{ij} - m_{ij}) \dot{z}_{j} - \sum_{j=1}^{k}\dot{m}_{ij}z_{j} +\dot{u}_i+w_i  \ne
&& -\varpi\mathrm{sgn}\left(u_{i} -\sum_{j=1}^{k} m_{ij}z_{j}\right)
-\frac{\rho}{e^*}\left(u_{i} -\sum_{j=1}^{k} m_{ij}z_{j}\right)  -\frac{\kappa}{(e^*)^{\gamma}}\left(u_{i}-\sum_{j=1}^{k} m_{ij}z_{j}\right)^{\gamma}
\label{m.twophase.fij}
\eea
where $z_{i}$ represents the state of the $i$th neuron,
corresponding to the $i$th element of $z(t)$,
$m_{ij}$ and $\dot{m}_{ij}$ denote respectively the $ij$th entries of $M(t)$ and $\dot{M}(t)$,
$u_i$ and $\dot{u}_i$ are the $i$th entries of $u(t)$ and $\dot{u}(t)$, respectively, and
$I_{ij}$ represents the $ij$th entry of the identity matrix $I$.

With \re{s.sgn}, the uncertainty $\bf{w}$ can be fully-rejected.
It is well known that the chattering phenomenon may occur, due to the $\mathrm{sgn}(\cdot)$ function is involved.
However, chattering
is highly undesirable in the implementation.
To further improve the computing performance and reject the impact of ${\bf w}$,
we adopt the following smooth compensation:
\bea  \label{ra.w}
s_i(e_i) = - \frac{\varpi^{2}e_i}{\varpi|e_i|+\varepsilon}
\eea
with $\varepsilon>0$.

\begin{theorem}
\label{theoremqp}
With rectifying action \re{rqp} or \re{sys.rhope} (with $\rho>0$), if the uncertainty compensation \re{ra.w} is adopted,
then as time increases,
the error $e_i(t)$ of ERS \re{err.dyn} will converge in finite time to a small residual set with the radius $\Delta$,
which can be adjusted by the design parameters. In particular,

i) as for rectifying action \re{rqp},
the residual set is established as
\bea   \label{rqp.bound}
\Delta = \Bigg{\{}e_i\Big{|} |e_{i}| \leq
\min \left\{\frac{2\varepsilon}{\rho''},\left(\frac{2\varepsilon}{\kappa_1''}\right)^{\frac{1}{\gamma_1}},
\left(\frac{2\varepsilon}{\kappa_2''}\right)^{\frac{1}{\gamma_2}} \right\}  \Bigg{\}};
\eea

ii) for rectifying action \re{sys.rhope}
with the exponent \re{sys.rhope}, \re{fda.qp.gamma} and \re{fda.qp.gamma-1},
the residual set is respectively calculated to be
\bea
\label{fastqp.bound}
 \Delta =
\Bigg{\{}e_i\Big{|} |e_{i}|\leq
e^{*} \min \left\{\frac{2\varepsilon}{\rho''},\left(\frac{2\varepsilon}{\kappa''}\right)^{\frac{1}{\gamma_1}},
\left(\frac{2\varepsilon}{\kappa''}\right)^{\frac{1}{\gamma_2}} \right\}  \Bigg{\}};
\eea

iii) with the exponent \re{gamma},
the resultant residual set $\Delta$ is
\bea   \label{fastqfe.bound}
\Delta = \Bigg{\{}e_i\Big{|} |e_{i}|\leq
e^{*} \min \left\{\frac{2\varepsilon}{\rho''},\left(\frac{2\varepsilon}{\kappa''}\right)^{\frac{2}{\beta}},
\left(\frac{2\varepsilon}{\kappa''e^{\frac{-\beta}{2m {\rm e}}}}\right)^{\frac{1}{\alpha}} \right\}  \Bigg{\}}
\eea
with parameters $\rho''>0, \kappa_1''>0, $ and $\kappa_2''>0$ to be specified.
\end{theorem}

{\bf Proof.} The proof is conducted for rectifying actions \re{rqp} and
 \re{sys.rhope} (with $\rho>0$), respectively.

i) For applying rectifying action \re{rqp},
let us choose the Lyapunov function candidate $V(e_i) = e_i^2$ (we use $V_i$ for simple notation), according to Lyapunov's second method.
Calculating the derivative of $V_i$ with respect to time yields
\bea
\dot V_{i} &=& -2 \rho e_{i}^2
- 2\kappa_1 e_{i}^{\gamma_1+1} - 2 \kappa_2 e_{i}^{\gamma_2+1} - 2 \frac{\varpi^{2} e_i^2}{\varpi |e_i|+\varepsilon}
+ 2e_i w_i   \ne
&\leq& -2\rho V_{i} -2\kappa_{1} V_{i}^{\frac{\gamma_1+1}{2}}
          -2 \kappa_{2} V_{i}^{\frac{\gamma_2+1}{2}}                 -2\frac{\varpi^{2}V_i}{\varpi \sqrt{V_i}
+\varepsilon}\ne
&&   +2\sqrt{V_i} \varpi \ne
&=&   -2\rho V_{i}   -2 \kappa_{1}  V_{i}^{\frac{\gamma_1+1}{2}}
          -2 \kappa_{2} V_{i}^{\frac{\gamma_2+1}{2}} +2\frac{\varpi \sqrt{V_i}}{\varpi \sqrt{V_i} +\varepsilon}\varepsilon \ne
&\leq&  -2\rho V_{i}
-2 \kappa_{1}  V_{i}^{\frac{\gamma_1+1}{2}}
-2 \kappa_{2} V_{i}^{\frac{\gamma_2+1}{2}}
+ 2\varepsilon \nn
\eea
Defining $\Lambda_{i}=\sqrt{V_i}$ leads to,
as $V_{i} \neq 0$,
\bea
\dot{\Lambda}_{i}&\leq&
-\rho \Lambda_{i}
-\kappa_{1} \Lambda_{i}^{\gamma_1}
-\kappa_{2} \Lambda_{i}^{\gamma_2}
+2\varepsilon  \nn
\eea

According to Lemma \ref{lem.RAtool},
let us choose $\rho=\rho'+\rho''$,
$\rho', \rho''>0$, such that
\bea  \label{DLam.t1}
\dot{\Lambda}_{i}&\leq&
- \rho' \Lambda_{i}
-\kappa_{1} \Lambda_{i}^{\gamma_1}
-\kappa_{2} \Lambda_{i}^{\gamma_2}
- \rho'' \Lambda_{i}
+2\varepsilon
\eea
It follows from \re{DLam.t1} that
as $\Lambda_i>2\varepsilon/\rho''$, $\dot{\Lambda}_{i} \leq-\rho' \Lambda_{i}-\kappa_{1}\Lambda_{i}^{\gamma_1} -\kappa_{2} \Lambda_{i}^{\gamma_2}$,
implying that $\Lambda_{i}$ converges to the set $\{\Lambda_{i}|\Lambda_{i} \le 2\varepsilon/\rho''\}$, and the convergence time can be estimated by \re{rqp.tg}.

Similarly, choosing parameters
$\kappa_1 = \kappa_1'+\kappa_1''$ and $\kappa_{2} = \kappa_{2}'+\kappa_2''$, in \re{DLam.t1},
respectively, we have
\bea
\dot{\Lambda}_{i}&\leq&
- \rho \Lambda_{i}
-\kappa_{1}' \Lambda_{i}^{\gamma_1}
-\kappa_{2} \Lambda_{i}^{\gamma_2}
- \kappa_{1}'' \Lambda_{i}^{\gamma_1}
+2\varepsilon   \nn
\eea
and
\bea
\dot{\Lambda}_{i}&\leq&
- \rho \Lambda_{i}
-\kappa_{1} \Lambda_{i}^{\gamma_1}
-\kappa_{2}' \Lambda_{i}^{\gamma_2}
- \kappa_{2}'' \Lambda_{i}^{\gamma_2}
+2\varepsilon   \nn
\eea
Then
the convergence bounds of $\Lambda_{i}$ can be estimated as
$\{\Lambda_{i}|\Lambda_{i} \le (2\varepsilon/\kappa_1'')^{1/\gamma_1}\}$,
and
$\{\Lambda_{i}|\Lambda_{i} \le (2\varepsilon/\kappa_2'')^{1/\gamma_2}\}$, respectively.

Therefore, considering the above three situations, the error $e_i$ will converge to the bound given in \re{rqp.bound},
and the estimate of the convergence time
corresponding to the bound
can be given by Theorem \ref{thm.rqp.ts}, with the chosen parameters.

ii) As for the exponents \re{sys.rhope}, \re{fda.qp.gamma} and \re{fda.qp.gamma-1},
the robust performance can be established on the basis of Theorems
\ref{thm.gammaC}-\ref{lem.fda.fastqp}.
The derivations are similar to those lines for case i).

iii) By adopting the exponent \re{gamma} in the rectifying action \re{sys.rhope} (with $\rho>0$), the derivative of $V$ can be calculated as
\bea
\dot V_{i} &=& -2\frac{\rho}{e^{\ast}} e_{i}^2
-2 \frac{\kappa}{(e^{\ast})^{\gamma}} e_{i}^{\gamma+1}  - 2\frac{\varpi^{2}e_i^2}{\varpi |e_i|+\varepsilon}
+ 2e_i w_i   \ne
&\leq& -2\frac{\rho}{e^{\ast}} V_{i} -2  \frac{\kappa}{(e^{\ast})^{\gamma}} V_{i}^{\frac{\gamma+1}{2}}
-2\frac{\varpi^{2}V_i}{\varpi \sqrt{V_i}
+\varepsilon}
 + 2\sqrt{V_i} \varpi \ne
&=&   -2\frac{\rho}{e^{\ast}} V_{i}   -2  \frac{\kappa}{(e^{\ast})^{\gamma}}  V_{i}^{\frac{\gamma+1}{2}}
           +2\frac{\varpi \sqrt{V_i}}{\varpi \sqrt{V_i} +\varepsilon}\varepsilon \ne
&\leq&  -2\frac{\rho}{e^{\ast}} V_{i}
-2  \frac{\kappa}{(e^{\ast})^{\gamma}}  V_{i}^{\frac{\gamma+1}{2}}
+ 2\varepsilon \nn
\eea
Defining $\Lambda_{i} = \sqrt{V_{i}}$ and $\Lambda^{\ast}=e^{\ast}$ leads to, as $V_{i} \neq 0$,
\bea
\dot \Lambda_{i} \leq
- \rho\frac{\Lambda_{i}}{\Lambda^{\ast}}
- \kappa \left(\frac{\Lambda_{i}}{\Lambda^{\ast}}\right)^{\gamma}+2\varepsilon , \nn
\eea

According to Lemma \ref{lem.RAtool},
let us choose $\rho = \rho'+\rho''$, $\rho', \rho''>0$.
In the case of  $\left(\Lambda_{i}/\Lambda^{\ast}\right) > (\eta/\rho'')$, we have
\bea
\dot \Lambda_{i} &\leq&
-\rho' \frac{\Lambda_{i}}{\Lambda^{\ast}}
- \rho''\frac{\Lambda_{i}}{\Lambda^{\ast}}
- \kappa \left(\frac{\Lambda_{i}}{\Lambda^{\ast}}\right)^{\gamma}+2\varepsilon  \ne
&\leq& -\rho' \frac{\Lambda_{i}}{ \Lambda^{\ast}}- \kappa\left(\frac{\Lambda_{i}}{ \Lambda^{\ast}}\right)^{\gamma} \nn
\eea
By Theorem \ref{theorem.fastqfe},
the estimation for the convergence time
can be made, respectively,
as $e_i$ approaches to
$\{\Lambda_{i}|\Lambda_{i} \le \Lambda^{\ast}(2\varepsilon/\rho'')\}$.

With the selection of $\kappa=\kappa'+\kappa''$, $\kappa', \kappa''>0$,
whenever $\left(\Lambda_{i}/\Lambda^{\ast}\right)^{\gamma} > 2 \varepsilon/\kappa''$,
 we obtain
 \bea
\dot \Lambda_{i} &\leq&
-\rho \frac{\Lambda_{i}}{\Lambda^{\ast}}
- \kappa' \left(\frac{\Lambda_{i}}{\Lambda^{\ast}}\right)^{\gamma}
- \kappa'' \left(\frac{\Lambda_{i}}{\Lambda^{\ast}}\right)^{\gamma}
+2\varepsilon  \ne
&\leq& -\rho \frac{\Lambda_{i}}{ \Lambda^{\ast}}- \kappa'\left(\frac{\Lambda_{i}}{ \Lambda^{\ast}}\right)^{\gamma}
\eea
By Theorem  \ref{theorem.fastqfe},
the convergence time can be
evaluated,
as $e_i$ achieves
the residual set given by
$\{\Lambda_{i}|\Lambda_{i} \le \Lambda^{\ast}(2\varepsilon/\kappa'')^{2/\beta}\}$ or
$\{\Lambda_{i}|\Lambda_{i} \le \Lambda^{\ast}(2\varepsilon/\kappa''e^{\frac{-\beta}{2m {\rm e}}})^{1/\alpha}\}$.
\QED

Theorem \ref{theoremqp} characterizes
the impact of design factors on
convergence performance of the constructed ERSs,
by which we can appropriately choose the design
factors, in order
to effectively improve the convergence rate
and adjust the residual set.
The radius of the residual set
depends on the adjustable constant $\varepsilon$, by which the computing error can be reduced dramatically,
as one select $\varepsilon$ to be small enough.
To appropriately choose design parameters $\rho''$ and $\kappa''$ can flexibly improve the convergence performance including the convergence time and the computing accuracy.
As such,
the computing error $e_i$ converges to
the residual set,
while the fast convergence rate is assured.
It is observed that
the convergence performance heavily depends on
the transition state $e^*$ introduced in rectifying actions.
By reducing $e^*$,
the convergence rate of the ERS can be sped up,
and the small radius of the residual set can be achieved.
In addition,
the uncertainty compensation \re{ra.w} is applicable for the case without the proportional term, i.e.
the rectifying action \re{qp}
and the rectifying action \re{sys.rhope} with $\rho=0$.
The robust performance of the constructed RNNs can be established
similarly to Theorem \ref{theoremqp}.

The proposed models have been applied and compared through numerical results, for time-variant matrix inversion, linear equation solving,
time-variant QP problem solving,
and repeatable motion planning of redundant manipulators,
in order to demonstrate effectiveness (involving
convergence and
robustness perofrmance)
of the proposed computing schemes.

\section{Conclusion}
\label{conclusion}
In this paper,
we present the error-recurrence approach to the time-variant problem solving
under uncertainty, by building into the constructive error dynamics of TZNNs the flexibility which is capable of enhancing robustness properly.
The ERS approach offering a control-theoretic methodology and the advantage lies in its simplicity in TZNN designs
and
their finite-duration stabilization designs
in a unified manner dealing with uncertainties, without assuming a specific type.
Novel rectifying actions have been constructed  to achieve the logarithmic settling-time and to speed up the convergence rate.
It is further proved that the settling time can be smaller
than the existing ones,
and the least upper bounds of the settling
time function are derived.
The theoretical results show that
the presented TZNN models not only possess global fixed-time convergence, but also provide
the uncertainty compensation to enhance
robustness with respect to uncertainties.
The applicability and effectiveness of the proposed
computing schemes have been verified through
solving time-variant matrix problems
and time-variant quadratic programming.

\end{document}